\documentclass[12pt,preprint]{aastex}

\shorttitle{Parsec-scale Rotation Measures in FRI galaxies}
\shortauthors{Kharb et al.}

\begin{document}

\title{Rotation-Measures across Parsec-scale Jets of FRI radio galaxies}

\author{P. Kharb\altaffilmark{}}
\affil{Department of Physics, Purdue University, West Lafayette, IN 47907}
\email{pkharb@physics.purdue.edu}
\author{D. C. Gabuzda\altaffilmark{}}
\affil{University College Cork, Cork, Ireland}
\author{C. P. O'Dea\altaffilmark{}}
\affil{Dept. of Physics, Rochester Institute of Technology, Rochester, NY 14623}
\author{P. Shastri\altaffilmark{}}
\affil{Indian Institute of Astrophysics, Koramangala, Bangalore 560034, India}
\and
\author{S. A. Baum\altaffilmark{}}
\affil{Center for Imaging Science, Rochester Institute of Technology, 
Rochester, NY 14623}

\begin{abstract}
We present the results of a parsec-scale polarization study of 
three FRI radio galaxies -- 3C66B, 3C78 and 3C264 -- obtained with the 
Very Long Baseline Array at 5, 8 and 15 GHz. Parsec-scale polarization has
been detected in a large number of beamed radio-loud active galactic nuclei,
but in only a handful of the relatively unbeamed radio galaxies.
We report here the detection of parsec-scale polarization at one or more 
frequencies in all three FRI galaxies studied. We detect Faraday rotation 
measures of the order of a few hundred rad~m$^{-2}$ in the nuclear jet 
regions of 3C78 and 3C264. In 3C66B polarization was detected at 8~GHz only. 
A transverse rotation measure gradient
is observed across the jet of 3C78. The inner-jet magnetic 
field, corrected for Faraday rotation, is found to be aligned along 
the jet in both 3C78 and 3C264, although the field becomes orthogonal further 
from the core in 3C78. The {\it RM} values in 3C78 and 3C264 are similar 
to those previously observed in nearby radio galaxies.
The transverse {\it RM} gradient in 3C78, the increase in the degree of 
polarization at the jet edge, the large rotation in the polarization angles 
due to Faraday rotation and the low depolarization between frequencies, 
suggests that a layer surrounding the jet with a sufficient number of thermal
electrons and threaded by a toroidal or helical magnetic field is a good
candidate for the Faraday rotating medium. This suggestion is tentatively
supported by {\it Hubble Space Telescope} optical polarimetry but
needs to be examined in a greater number of sources. 
\end{abstract}


\section{Introduction}

Knowledge of the ambient medium near the jet launching site in the central 
parsecs of an active galactic nucleus (AGN) can shed light on why these 
AGNs produce jets with different morphologies: low-luminosity Fanaroff-Riley 
type I radio galaxies \citep[FRI;][]{FanaroffRiley74} have jets that flare into 
diffuse radio lobes on scales of a few hundred parsec, 
while the narrow, collimated jets in the high-luminosity FR type II radio 
galaxies terminate in bright hot spots and create radio lobes via plasma 
backflow. Although the origin of the FR dichotomy is an open question, and a 
number of differences in the two classes have been cited as potential
causes, differences in the nuclear environments of FRIs and FRIIs, which in 
turn influence the evolution of the jet due to interaction and entrainment, are 
widely believed to be important
\citep[e.g.,][]{Begelman82,Baum95,Bicknell95,DeYoung96,Laing96}.

Very Long Baseline Interferometry (VLBI) is currently the only technique by 
which the parsec-scale environments of radio galaxies can be probed. 
Multi-frequency polarization sensitive VLBI experiments can determine the
Faraday rotation measure ({\it RM}) arising in the vicinity of
the parsec-scale radio jets.
Radiation propagating through a magneto-ionic medium has its plane of
polarization rotated from its intrinsic value $\chi_0$ to $\chi$ 
following the relation
\begin{equation}
\chi=\chi_0 + RM\lambda^2
\end{equation} where 
\begin{equation}
RM= 812 \int n_e B_{\|} dl~~~\mathrm{rad~m^2}
\end{equation}
$n_e$ being the electron density in cm$^{-3}$, $B_{\|}$ the net 
line of sight magnetic field in milliGauss, and $dl$ an element of path 
length through the plasma in parsecs \citep[see review by][]{Gardner66}. 
The net parsec-scale {\it RM} is dominated by contributions from the nuclear
environment of the AGN in its host galaxy and our own Galaxy; the
Galactic contribution is usually estimated using arcsecond-resolution
observations at multiple wavelengths near 18--20~cm 
\citep[e.g.,][]{RudnickJones83}.
VLBI rotation measures have been observed and studied in a number of
relativistically beamed AGNs, $i.e.,$ radio-loud quasars and BL~Lac objects. 
Such studies have indicated that contributions to the {\it RM} from regions 
close to the radio jets are dominant
%and that fundamental environmental differences exist between quasars and BL~Lacs 
\citep[e.g.,][]{ZavalaTaylor03,ZavalaTaylor04}.

Radio galaxies are believed to be the plane-of-sky counterparts of quasars 
and BL~Lacs \citep{UrryPadovani95}. Since their cores and jets are not as 
strongly Doppler boosted and are therefore radio faint, there have been 
relatively few VLBI polarization studies done on them, and polarization has 
been detected on parsec-scales in only a handful
\citep{Taylor01,Middelberg04,ZavalaTaylor02,Kharb05,TaylorGugliucci06}.
Parsec-scale rotation measures have only been estimated for the 
FRII radio galaxies 3C111, 3C120 and 3C166, and the FRI radio 
galaxies M87 and 3C84. 
Here we augment these data with the parsec-scale polarization images of 
three FRI radio galaxies $-$
3C66B, 3C78 and 3C264, along with the {\it RM} images of the latter two. 

We have adopted a cosmology in which $H_0$=71 km s$^{-1}$ Mpc$^{-1}$, 
$\Omega_m$=0.27 and $\Omega_{\Lambda}$=0.73.
The spectral indices $\alpha$ are defined such that the flux density
$S_\nu$ at frequency $\nu$ is $S_\nu \propto \nu^{-\alpha}$.

\section{The FRI galaxies under study}

We observed four FRI radio galaxies from a larger sample of radio-loud sources 
\citep{Kharb04} that possessed optical nuclei in their {\it Hubble 
Space Telescope (HST)} images \citep{Chiaberge99,Chiaberge00,VerdoesKleijn02}, 
and had a correlated VLBI flux density $\geq$100~mJy, with global VLBI at 
8.4~GHz in March 2002. 
Parsec-scale polarized emission was detected in all of them \citep[][]{Kharb05}. 
New observations of three of the four FRIs that showed 
polarization in their parsec-scale jets, $viz.,$ 3C66B, 3C78 and 3C264, 
were obtained at 5.0, 8.4 and 15.3 GHz in September 2005 with a global 
VLBI array including the Very Long Baseline Array (VLBA) and the Effelsberg 
antenna, and are the focus of the present work. 

3C66B resides in an elliptical galaxy at a distance of 91 Mpc that forms 
a part of a dumbbell pair close to the Abell cluster 347. It is one of the 
few FRI radio galaxies in 
which the dominant radio jet has an infrared \citep{Tansley00}, 
optical \citep{Fraix-Burnet97}, and an X-ray counterpart 
\citep{Hardcastle01}. Kiloparsec-scale observations of 
3C66B suggest the jet inclination to be less than $\sim53\degr$ 
\citep{Hardcastle96}, while an inclination for the VLBI jet of $\sim45\degr$ 
has been inferred by \citet{Giovannini01}. X-ray observations have revealed 
the presence of copious amounts of hot gas, with which the kiloparsec radio jet seems 
to be interacting \citep{Croston03}. 

3C78 resides in the isolated, small E/S0 galaxy NGC1218 \citep{Matthews64}
at a distance of 124 Mpc. 
{\it HST} observations have revealed that the optical galaxy has a face-on dust 
disk and an optical jet \citep{Sparks95}. 
\citet{Baum88} have reported that the innermost emission-line gas is 
elongated roughly in the direction of the radio jet, following the
``alignment effect'' \citep[e.g.,][]{McCarthy87,Privon08}.

The radio morphology of 3C264 is of the narrow-angle-tail type 
\citep{ODeaOwen85}. It is hosted by an S0 galaxy 93 Mpc away that 
lies in a dense part of the Abell cluster 1367 \citep{Schombert86}.
Like 3C66B and 3C78, the radio jet in 3C264 also has an optical
counterpart \citep{Crane93}. {\it HST} images show that the optical jet is 
projected onto a face-on dust disk \citep{deKoff00}. 
A jet inclination angle of $\sim50\degr$ has been inferred on 
kiloparsec- \citep{Baum97,Baum98} and parsec-scales \citep{Giovannini01}.

\section{Observations and Data Reduction}

Observations were carried out with the ten elements of 
the VLBA and the 100-m Effelsberg telescope at
5.0, 8.4 and 15.3 GHz on 10 Sept 2005, for a total of 24 hours. The three radio
galaxies were each observed for $\approx$2.5 hours at each frequency.
3C84 was used as the instrumental polarization calibrator, assuming it to be
unpolarized at these frequencies. The compact AGN 1156+295 was used
as the electric vector position angle (EVPA or $\chi$) calibrator. Five to six
several-minute scans of each of these sources at each frequency were
interspersed throughout the observing schedule.

The data were reduced using standard procedures in the Astronomical Image 
Processing System (AIPS) software package, using Los Alamos as the reference 
antenna at all stages of the calibration. Amplitude calibration was
performed using the system temperatures and gain curve provided and the 
AIPS task APCAL, while the group delay and phase-rate calibration were 
performed using the AIPS task FRING. The cross polarized delays were 
determined using the procedure VLBACPOL. Finally the residual delays and
delay rates were derived using FRING. The instrumental polarization 
leakage terms were determined using 
the final self-calibrated map of 3C84 in the task LPCAL. 
These were typically of the order of $1-2\%$ at 5.0 and 
8.4 GHz and $1-3\%$ at 15.3 GHz for all the antennas. 

We determined the absolute EVPA calibration by comparing the total 
VLBI-scale polarization of 1156+295 with integrated polarization obtained 
from monitoring databases maintained by the NRAO 
\footnote{http://www.vla.nrao.edu/astro/calib/polar/} and the
University of Michigan Radio Astronomy Observatory (the latter data
kindly provided by H. and M. Aller). Unfortunately, there were no
integrated polarization measurements with good signal-to-noise ratio
available on the actual date of our VLBI experiment, making it necessary
to interpolate these measurements to our VLBA date. This was
straightforward for the 5~GHz EVPA calibration, since the NRAO database 
contained 5~GHz measurements on 4 and 26 September that
virtually exactly coincided (differing by only $1\degr$ in the polarization 
angle $\chi$, with errors in $\chi$ being $0.5-1.5\degr$). Accordingly,
we adopted the integrated 5~GHz EVPA for 1156+295 to be $-86\degr = 
+94\degr$, and consider the overall uncertainty in our 5~GHz EVPA 
calibration to be no larger than $1-2\degr$. 

NRAO measurements were also available at 8.4~GHz on these two dates, but 
showed a larger difference: $\chi = 81\degr\pm 1\degr$ on 4 September, and
$56\degr\pm 3\degr$ on 26 September. In addition, the UMRAO
database had an 8~GHz measurement on 24 September with $\chi = 
75\degr\pm 8\degr$; this latter measurement is of limited use
due to its large uncertainty, although it suggests that the EVPA of
1156+295 was not wildly variable near the time of our VLBA observations. 
If the $25\degr$ change between the integrated EVPAs for the two 
NRAO measurements was fairly smooth, the 8.4~GHz
EVPA on our observing date would have been $74\degr$, and we adopted
this as the integrated 8.4~GHz EVPA. This value is obviously somewhat
uncertain; we accordingly consider the overall uncertainty in our 8.4~GHz
EVPA calibration to be $3-4\degr$. 

The NRAO polarization monitoring database does not include measurements at or 
near 15.3~GHz; the UMRAO database had two measurements near our VLBA observing date,
but corresponding to very different $\chi$ values, each with relatively large 
uncertainties. Therefore, we explored two different ways to obtain a
estimate of the integrated 15.3~GHz $\chi$ for our VLBA date: 
(i) obtain a linear fit of polarization angle $\chi$ versus observing 
wavelength $\lambda^2$ for the 4 September 2005 NRAO data at 5, 8.4, and 22~GHz 
and use the resulting {\it RM} together with the 
reliably inferred 5~GHz integrated $\chi$ for 1156+295 on 10 September 
to derive a 15.3~GHz $\chi$ value for that date; (ii) interpolate
the NRAO 22~GHz $\chi$ values for 1156+295 for 4 September ($63\degr$)
and 26 September ($99\degr$) in the same way that we did for the 8.4~GHz $\chi$;
derive a $\chi$ versus $\lambda^2$ plot using the
interpolated 5, 8.4 and 22~GHz polarization angles for 10 September, and
use the resulting {\it RM} for 10 September together with 
the reliably inferred 5~GHz $\chi$ value for 1156+295 for 10 September 
to derive a 15.3~GHz $\chi$ value for that date. In both cases, the plots 
of $\chi$ versus $\lambda^2$ could be approximated by linear fits 
within the errors. Approach (i) yielded $\chi_{15.3} \approx
68\degr$, while approach (ii) yielded $\chi_{15.3} \approx 72\degr$. 
Realizing that some uncertainty in our EVPA calibration at 15~GHz was
inevitable given the limited calibration information, but encouraged by the 
fact that these two estimates were similar, we adopted the average value of
$70\degr$. We estimate the overall uncertainty in our 15.3~GHz
EVPA calibration to be $5-6\degr$.

{ 
For reference, a total EVPA uncertainty of $6^{\circ}$ between 5 
and 8~GHz translates into an uncertainty in the corresponding {\it RM} of about
45~rad\,m$^{-2}$.
It is important to note that the application of an incorrect
EVPA calibration will add the same offset to the observed EVPAs at
every pixel in the map in which polarization is detected. Therefore, this can
certainly bias the derived rotation measures, but cannot give rise to spurious
{\it RM} gradients. Thus, although the absolute rotation measures that are 
derived will have some added uncertainty due to the uncertainty in our EVPA 
calibration, this is not propagated to an added uncertainty in the {\it RM} gradients.}

The AIPS tasks IMAGR and CALIB were used iteratively for the imaging
and self-calibration, respectively. Stokes $Q$ and $U$ maps were also made using
IMAGR and the final, fully self-calibrated data. 
Pairs of the final Stokes $I$ images at neighboring frequencies (5 + 8~GHz and 
8 + 15~GHz) were made with identical beams, and used to derive spectral-index
maps using the task COMB. Since information about the absolute position of the 
VLBI structure is inevitably lost during the phase self-calibration, it is
necessary to first align the two $I$ maps.
The hybrid-mapping processes essentially aligns the VLBI cores, which is not
the correct absolute alignment, since the core position is frequency-dependent
\citep{Lobanov98}.
We determined the relative alignment between pairs of $I$ images based on
a comparison of their optically thin regions of emission using the
program of \citet{Croke08}; note that the algorithm used does
not rely on the presence of distinct components at each frequency, and
so is well suited to the images considered here.

The Stokes $Q$ and $U$ maps were used in the task COMB to obtain maps of the 
polarization intensity and polarization position angle. These 
polarization-angle maps at the different frequencies were used in either 
COMB or the task `RM' to obtain two-frequency or three-frequency 
rotation-measure maps, respectively. Prior to obtaining the {\it RM} maps, 
versions of the $Q$, $U$, and polarization-angle maps were made with identical 
beams. Polarization values with S/N less than 3 were blanked in COMB to 
produce the polarized intensity maps. The polarization angle maps were 
restricted (by blanking in COMB) to have output errors less than 
$\approx10\degr$ . The three-frequency {\it RM} map of 3C78 was made using 
the AIPS task `RM', with unblanked polarization angle maps and 
$\sigma_{RM}$ blanking in `RM' such that $\sigma_{RM}<55$~rad~m$^{-2}$. 
Estimates for the peak surface brightness and the $RMS$ noise in the 
final total and polarized intensity maps were obtained using AIPS 
procedures TVMAXF and TVWIN \& IMSTAT, respectively. Depolarization maps
were made using the degree of polarization maps at 5, 8 and 15 GHz for 3C78 
and 5 and 8 GHz for 3C264. The polarization intensity maps were first 
corrected for Ricean bias \citep[e.g.,][]{Killeen86} by using the 
AIPS task POLCO. 

\section{Results}

\begin{figure}[h]
\centering{
\includegraphics[width=8.1cm]{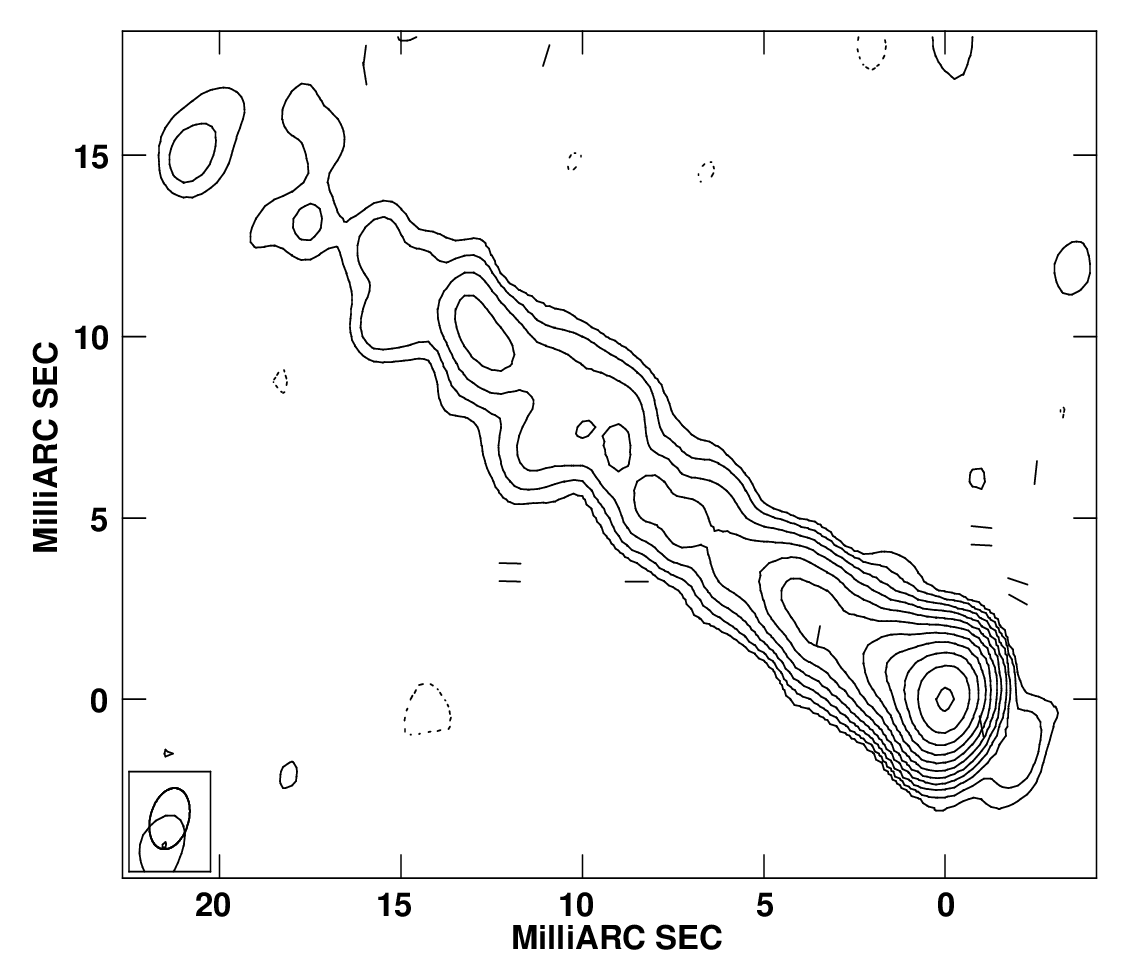}
\includegraphics[width=8.1cm]{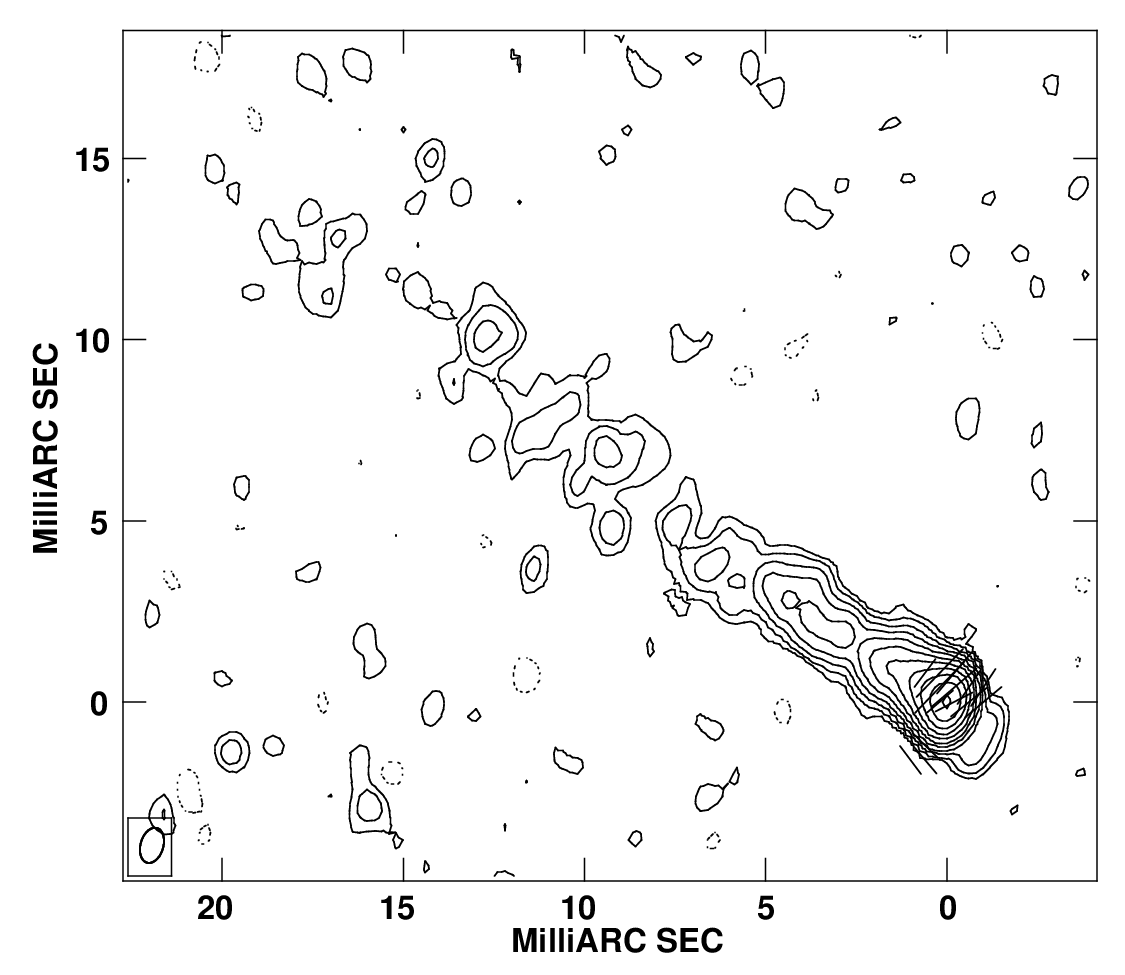}
\includegraphics[width=8.1cm]{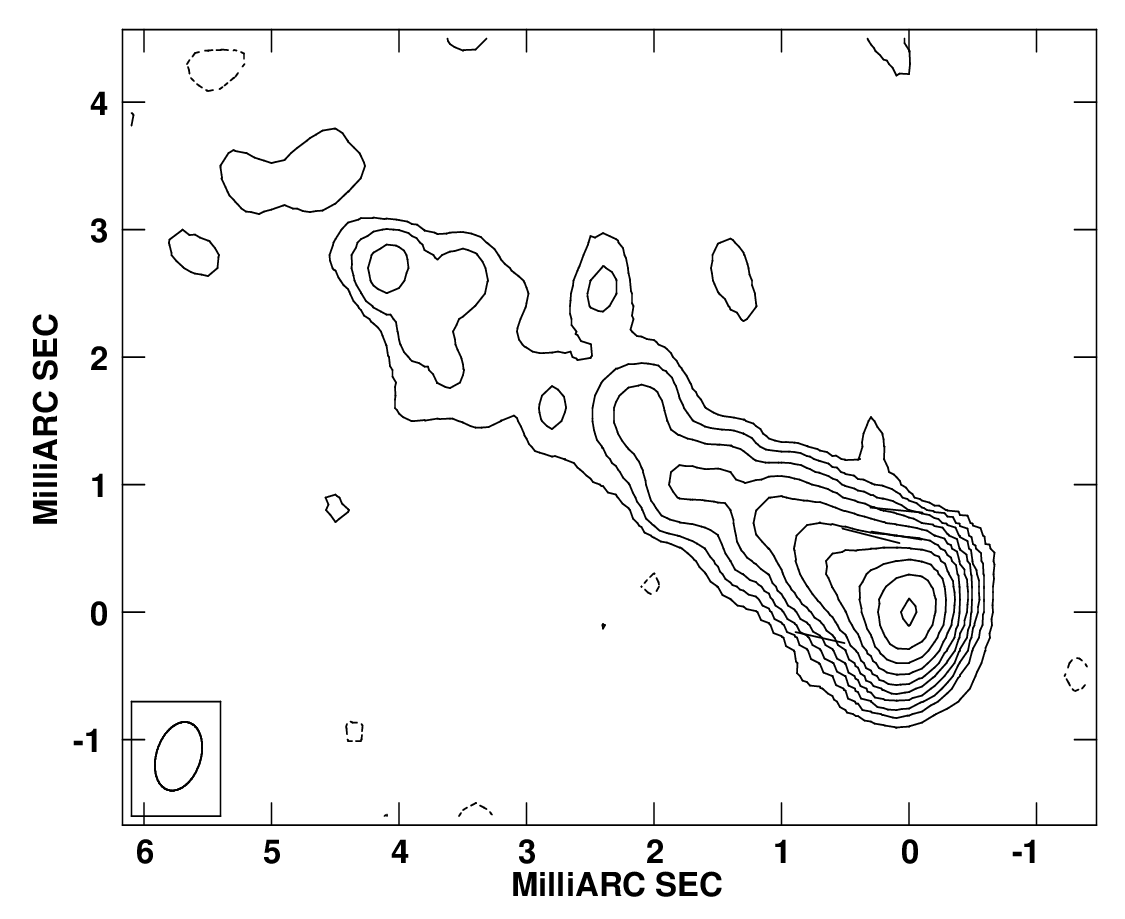}}
\caption{Total intensity maps of 3C66B at 5.0~GHz (top left), 8.4~GHz (top right) 
and 15.3 GHz (bottom) with polarization vectors superimposed. The contours are in 
percentage of the peak surface brightness and increase in steps of 2. The lowest
contour and peak brightness for 5, 8 and 15 GHz are $\pm$0.085\% of 
114~mJy~beam$^{-1}$, $\pm$0.085\% of 111~mJy~beam$^{-1}$, and $\pm$0.085\% of 
97~mJy~beam$^{-1}$, respectively. 
$\chi$ vectors for 8 GHz: 1 mas = 0.3 mJy beam$^{-1}$.}  
\label{fig:3c66b}
\end{figure}

\begin{figure}[h]
\centering{
\includegraphics[width=8.1cm]{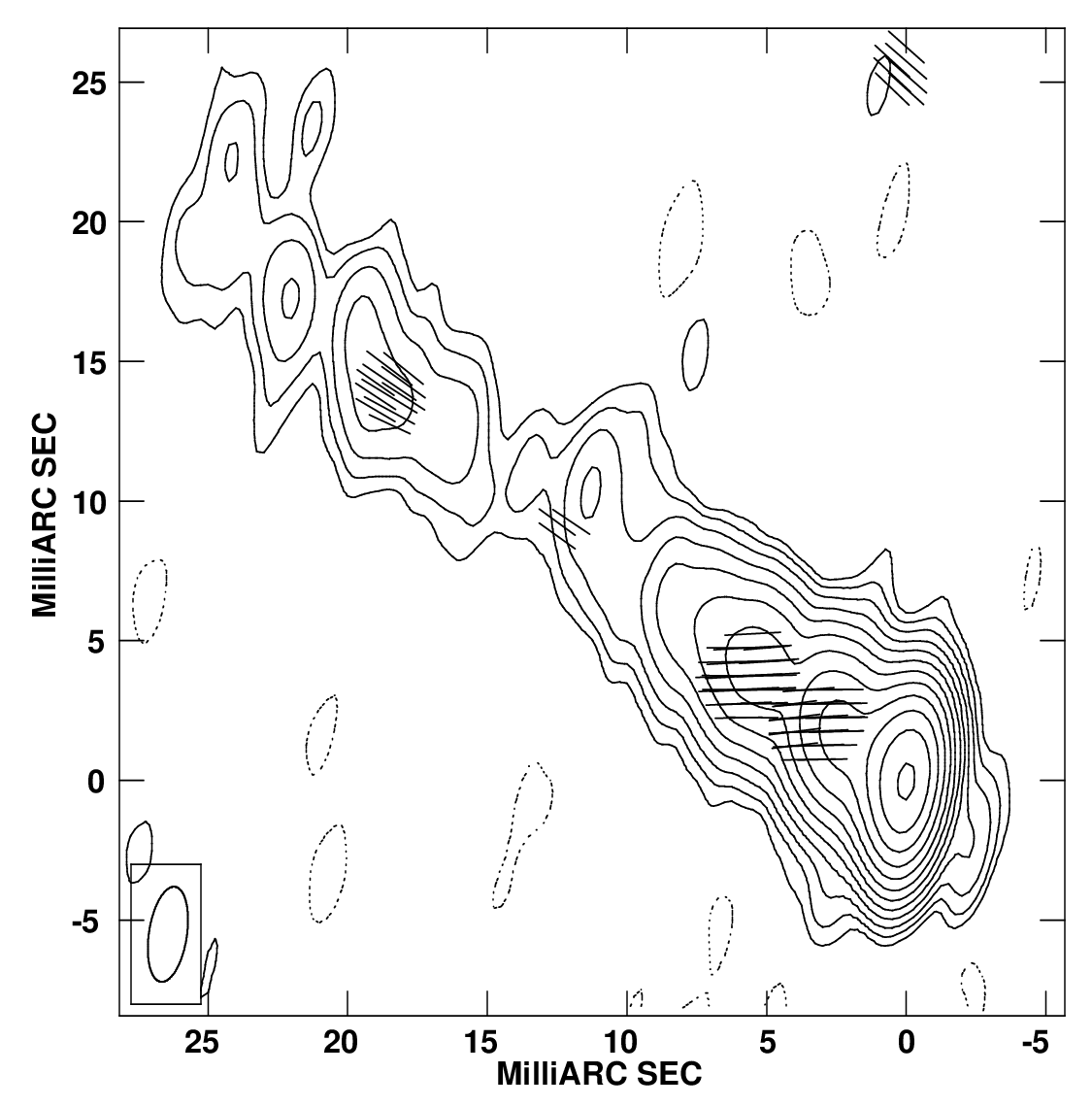}
\includegraphics[width=8.1cm]{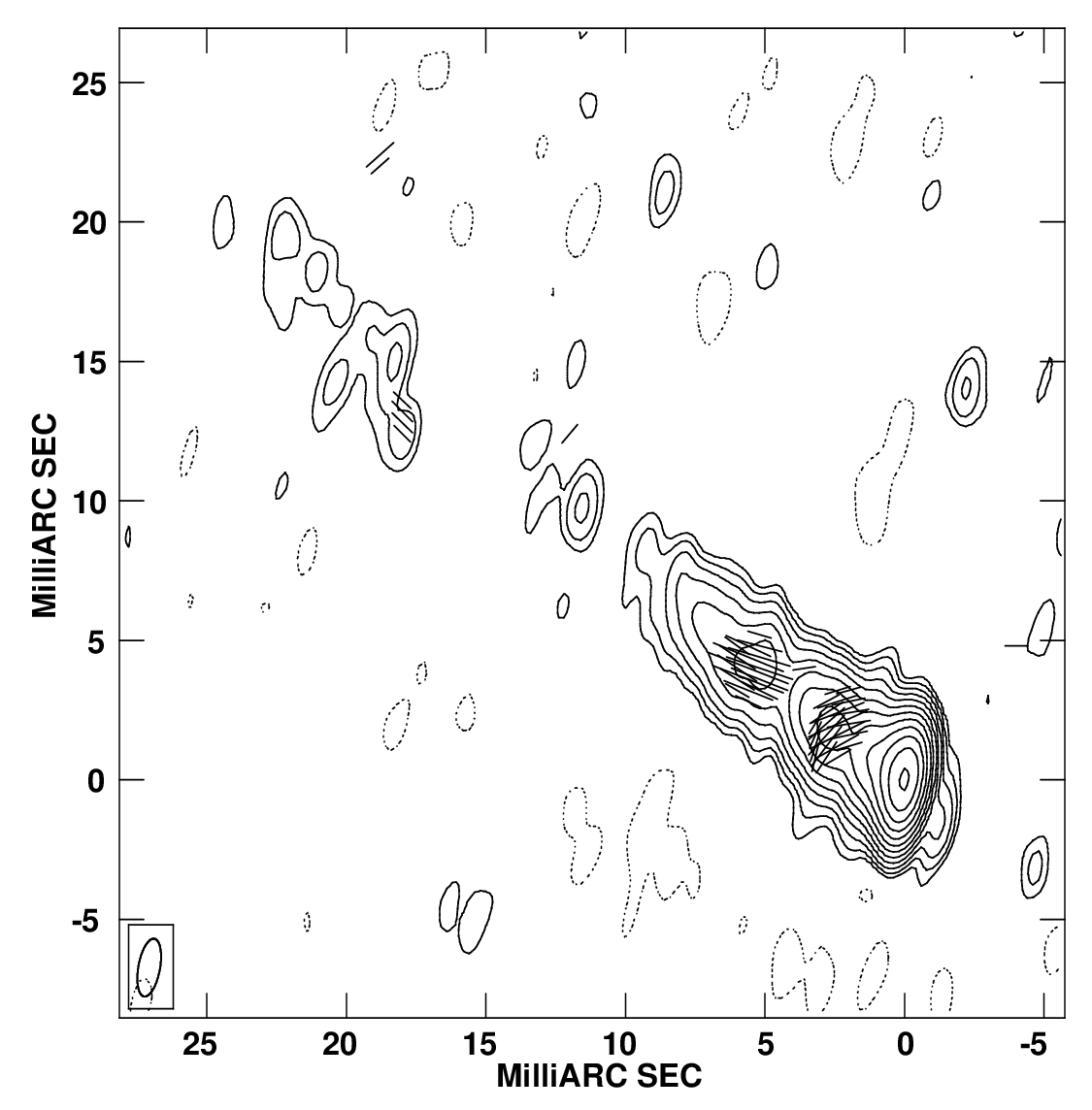}
\includegraphics[width=8.1cm]{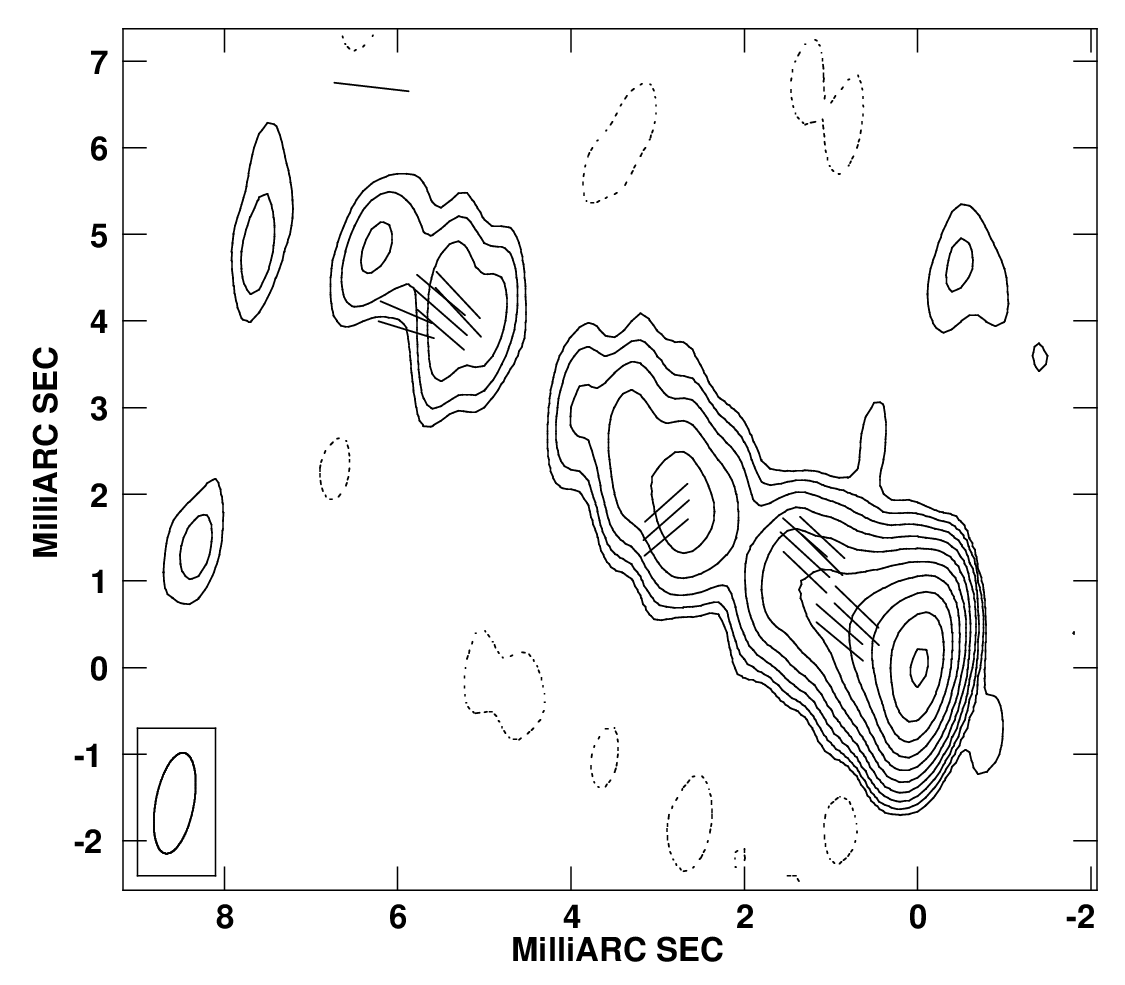}}
\caption{Total intensity maps of 3C78 at 5.0~GHz (top left), 8.4~GHz (top right)
and 15.3 GHz (bottom) with polarization vectors superimposed. The contours are in 
percentage of the peak surface brightness and increase in steps of 2. The lowest
contour and peak brightness for 5, 8 and 15 GHz are $\pm$0.042\% of 
342~mJy~beam$^{-1}$, $\pm$0.042\% of 331~mJy~beam$^{-1}$, and $\pm$0.17\% of 
242~mJybeam$^{-1}$, respectively. 
$\chi$ vectors for 5, 8 and 15 GHz: 1 mas = 0.25, 0.3, and 0.8 mJy beam$^{-1}$,
respectively.}
\label{fig:3c78}
\end{figure}

\begin{figure}[h]
\centering{
\includegraphics[width=6.4cm]{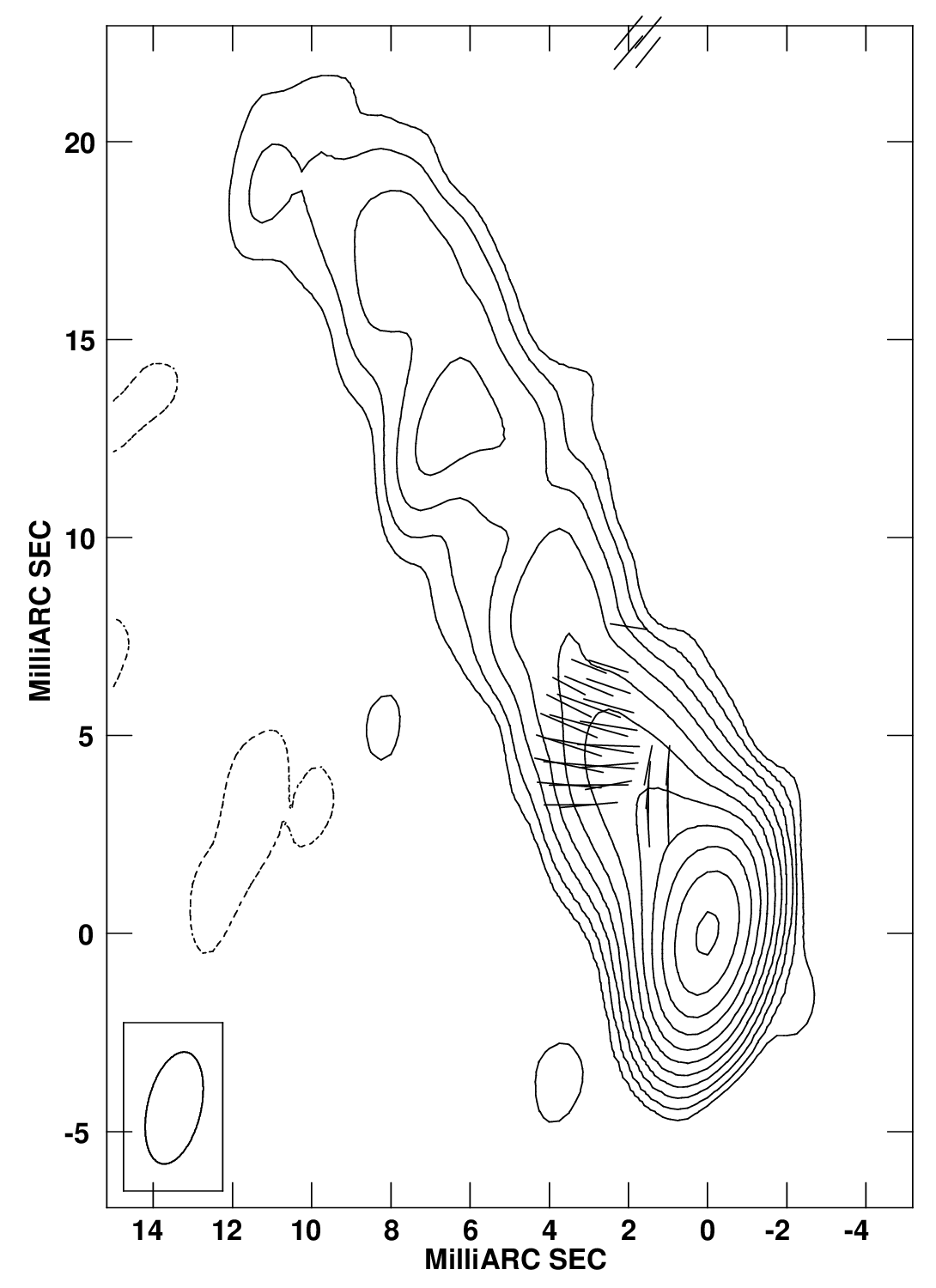}
\includegraphics[width=6.4cm]{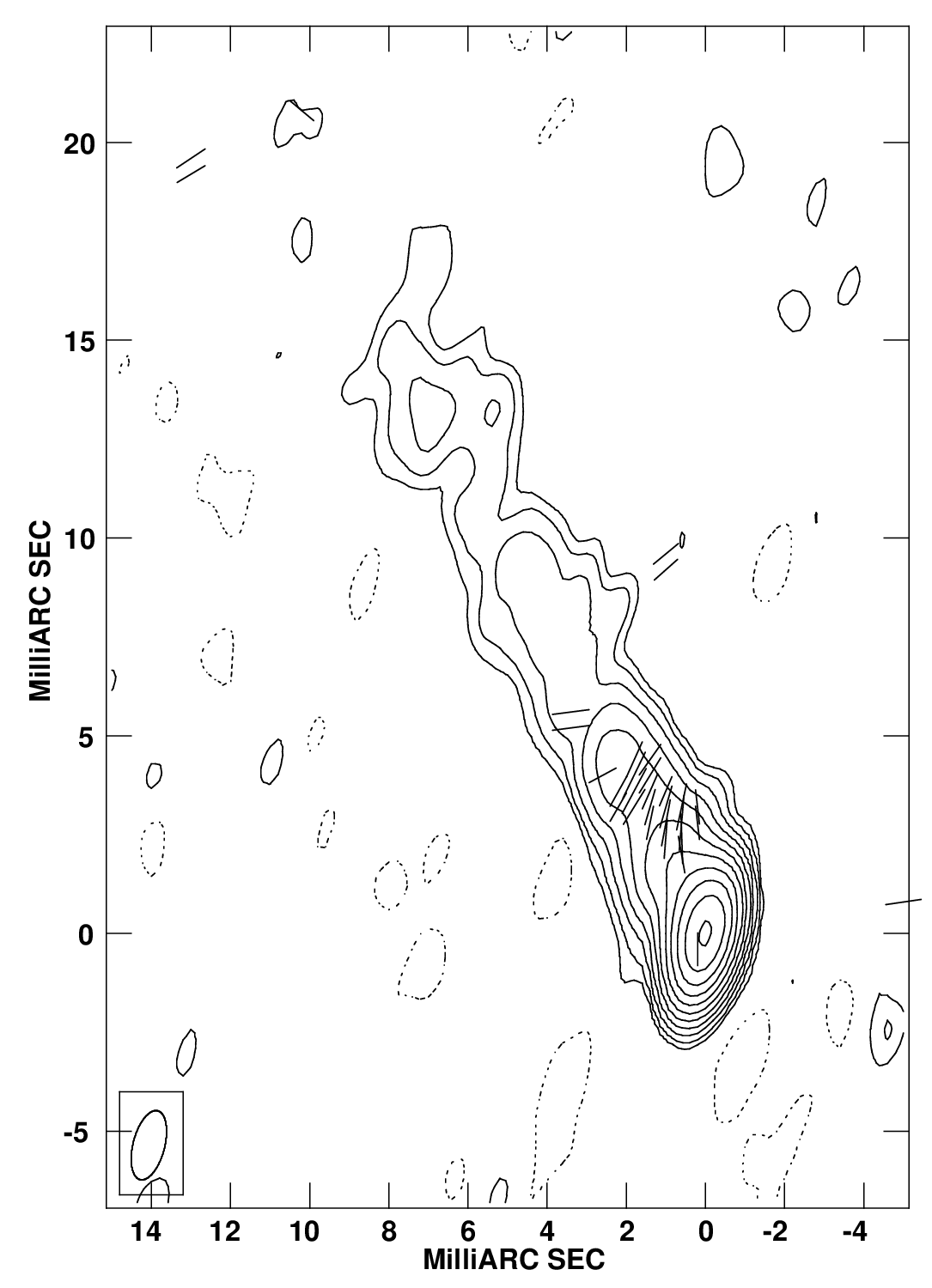}
\includegraphics[width=6.4cm]{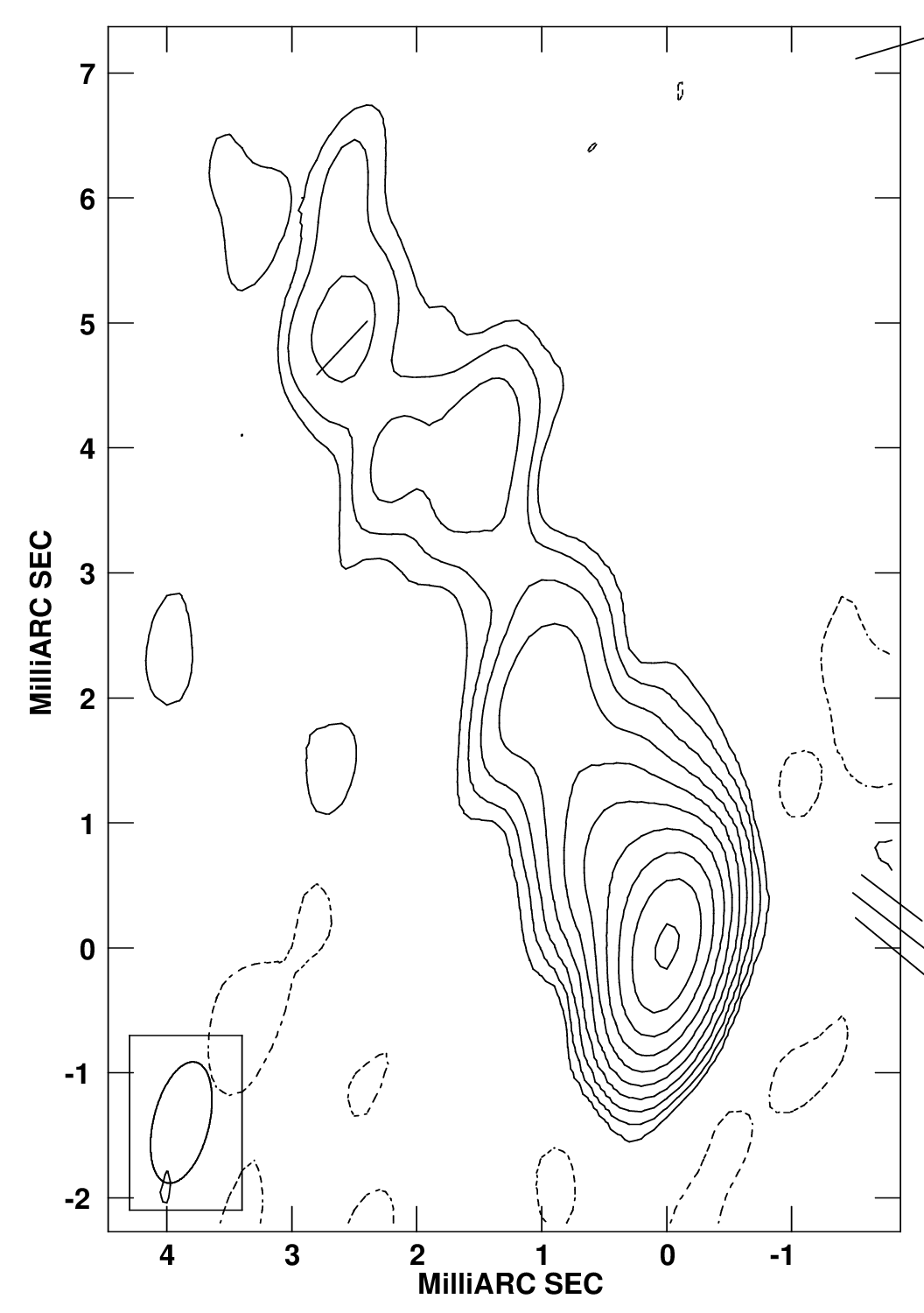}}
\caption{Total intensity maps of 3C264 at 5.0~GHz (top left), 8.4~GHz (top right)
and 15.3 GHz (bottom) with polarization vectors superimposed. The contours are 
in percentage of the peak surface brightness and increase in steps of 2. 
The lowest contour and peak brightness for 5, 8 and 15 GHz are $\pm$0.085\% of 
144~mJy~beam$^{-1}$, $\pm$0.085\% of 140~mJy~beam$^{-1}$, and  
$\pm$0.17\% of 106~mJy~beam$^{-1}$, respectively. $\chi$ vectors for 5 and 8 GHz: 
1 mas = 0.25 and 0.3 mJy beam$^{-1}$, respectively.}
\label{fig:3c264}
\end{figure}

\begin{figure}[h]
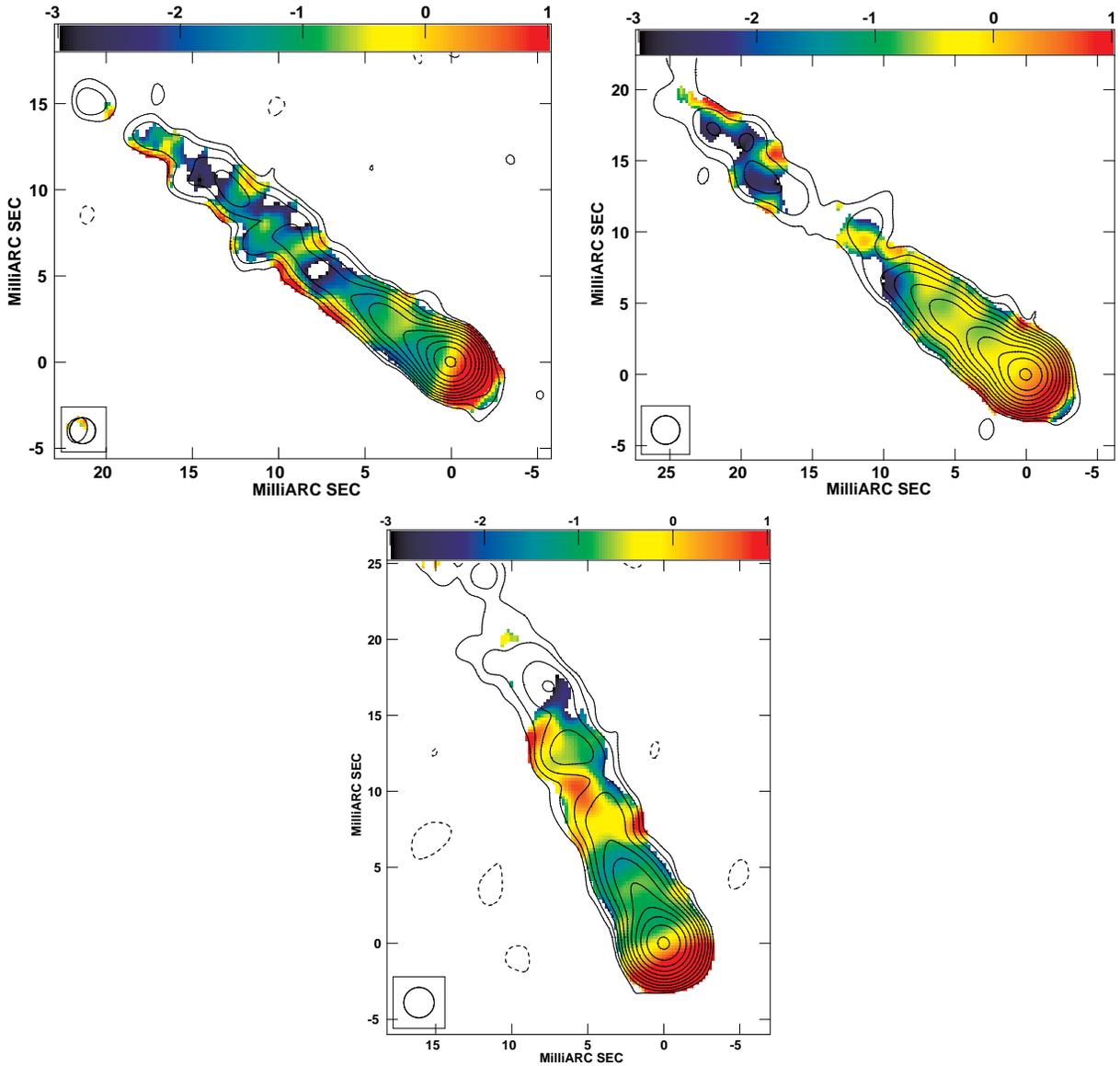

\centering{
\includegraphics[width=8.3cm]{f4a.ps}
\includegraphics[width=8.0cm]{f4b.ps}
\includegraphics[width=6.4cm]{f4c.ps}}
\caption{The 5$-$8~GHz spectral index map of (top left) 3C66B made with 
a 1.5~mas circular beam, (top right) 3C78 and (bottom) 3C264, each made with 
a 2~mas circular beam. The 5~GHz total intensity contours are superimposed. 
Note the optically thin counterjet emission in 3C66B and 3C78.}
\label{fig:spix}
\end{figure}

\begin{figure}[h]
\centering{
\includegraphics[width=17cm]{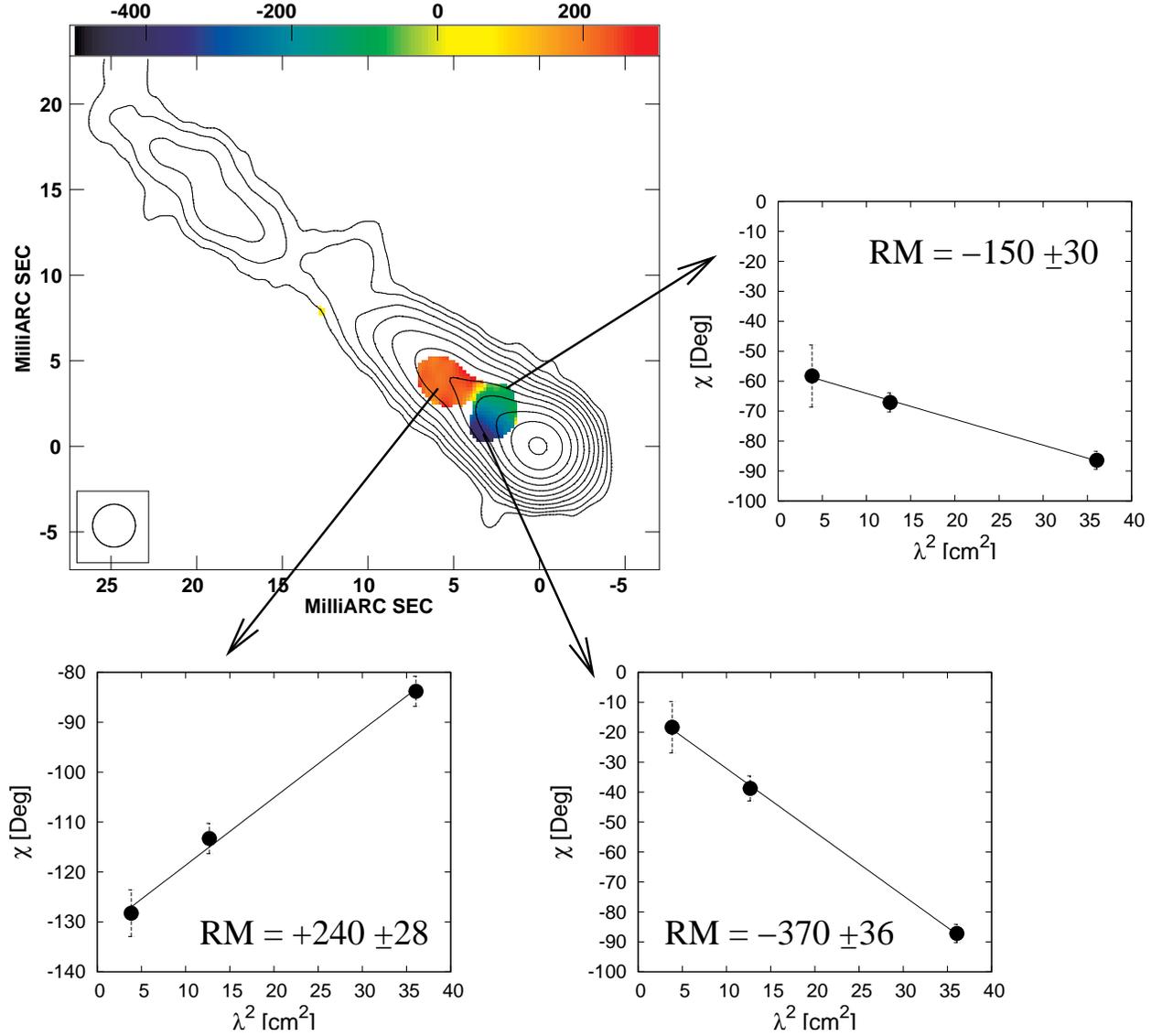}}
\caption{The 3-frequency rotation measure image of 3C78 with 5~GHz total 
intensity contours superimposed. An area covering four pixels (pixel 
size=0.2~mas) was used to create the inset $\chi$-$\lambda^2$ plots. The {\it RM}
values in the inset plots {and colour scale} are in units of rad~m$^{-2}$,
{with $\sigma_{RM}<55$~rad~m$^{-2}$.} 
Rotation measures in the rest-frame of the source are larger by a factor 
$(1+z)^2$ or 1.058.}
\label{fig:3c78rm}
\end{figure}

\begin{figure}[h]
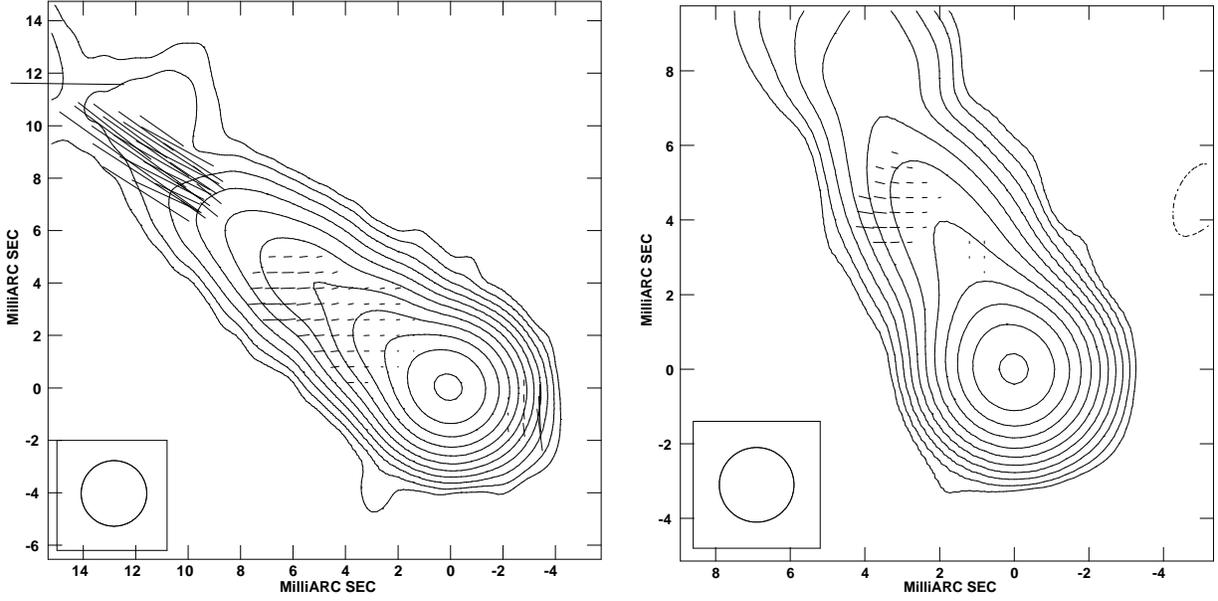

\centering{
\includegraphics[width=8.3cm]{f6a.ps}
\includegraphics[width=8.0cm]{f6b.ps}}
\caption{5~GHz total intensity contours with degree of polarization
vectors superimposed for 3C78 (left) and 3C264 (right).
1 mas ticks correspond to 8\% and 50\% for 3C78 and 3C264,
respectively. Degree of polarization clearly increases towards the southern
and eastern edge of the 3C78 and 3C264 jets, respectively.}
\label{fig:fpol}
\end{figure}

\begin{figure}[h]
\centering{
\includegraphics[width=17cm]{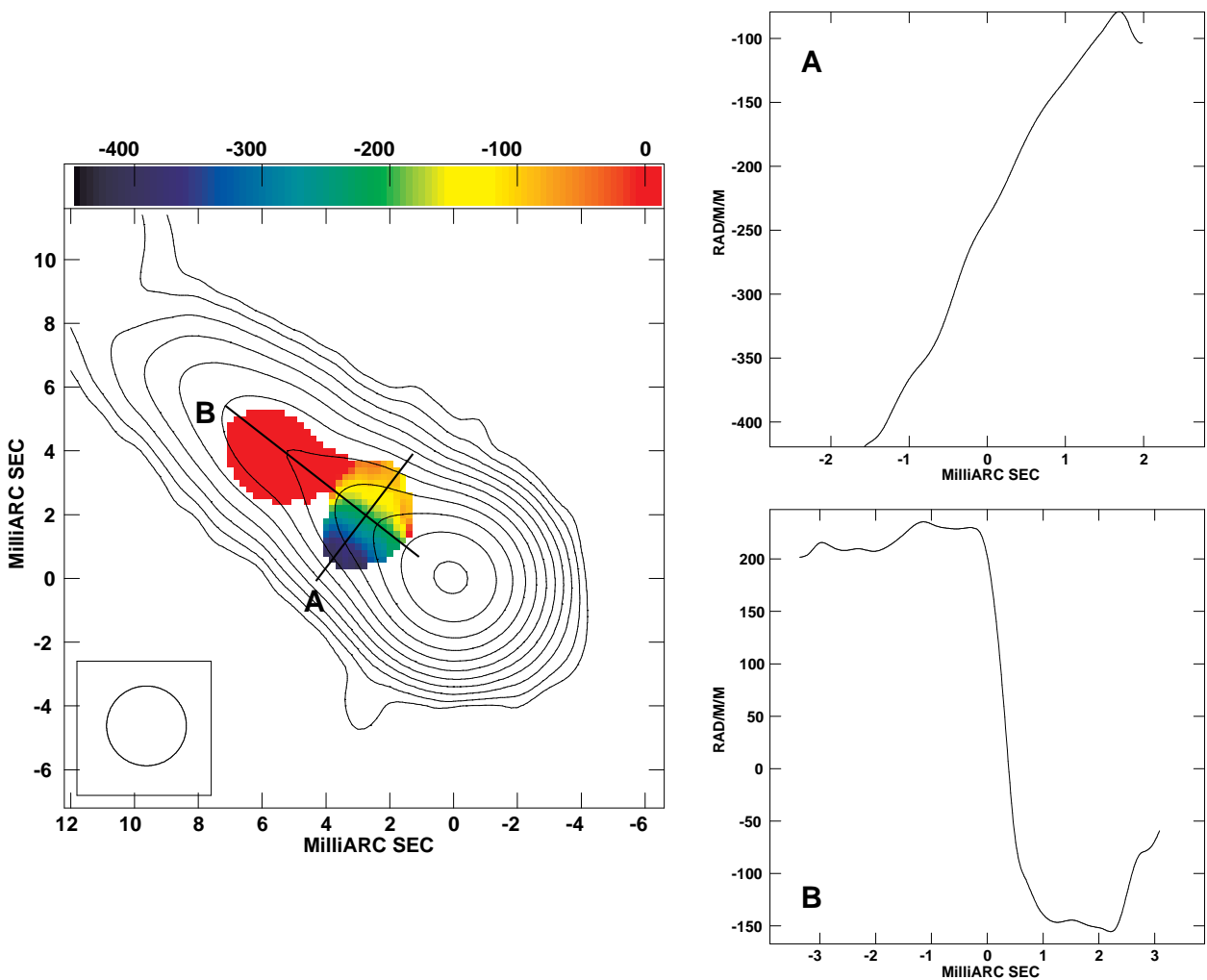}}
\caption{The {\it RM} gradient across the jet in 3C78. 
The left panel zooms in and displays a narrower range of {\it RM} where the
{\it RM} gradient is clearly visible perpendicular to the jet close to the core.
The {\it RM} changes from $\sim-$450 to $\sim-$50 rad~m$^{-2}$ over a region 5 mas 
(two beam-sizes) across {the jet (Inset A). Inset B displays the rotation
measure along the jet direction.}
Rotation measures in the rest-frame of the source are larger by a factor 
$(1+z)^2$ or 1.058.}
\label{fig:3c78sl}
\end{figure}

\begin{figure}[h]
\centering{
\includegraphics[width=12cm]{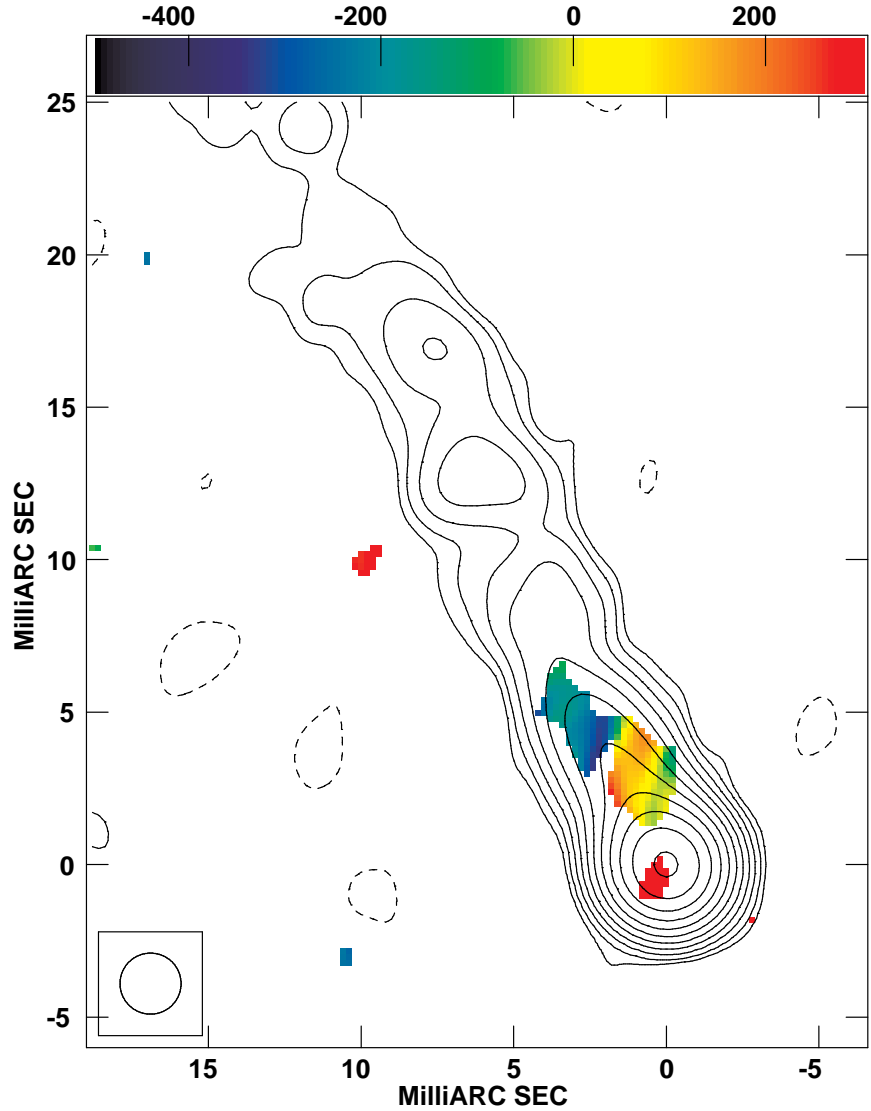}}
\caption{The rotation measure image of 3C264 with 5~GHz total intensity
contours superimposed. Only the 5 and 8~GHz images were used to derive
the {\it RM}. {The colour scale is in units of rad~m$^{-2}$ with
$\sigma_{RM}$ being typically $<50$~rad~m$^{-2}$}.
Rotation measures in the rest-frame of the source are larger by a factor 
$(1+z)^2$ or 1.044.}
\label{fig:3c264rm}
\end{figure}

The total intensity radio images with polarization electric vectors superimposed
are presented in Figs.~\ref{fig:3c66b}$-$\ref{fig:3c264}, the 5$-$8~GHz 
spectral-index maps in Fig.~\ref{fig:spix} and rotation-measure maps and plots 
in Figs.~\ref{fig:3c78rm}$-$\ref{fig:3c264rm}. 
Table~\ref{tabobsparam} lists some parameters pertaining to the final
images. We discuss our results for each radio galaxy below.


{\it 3C66B:}
Our observations reveal tentative polarization close to the radio core only 
at 8 GHz, consistent with the epoch-2002 observations of \citet{Kharb05}.  
We failed to detect any polarization at 5 and 15~GHz. 
\citet{Simard-Normandin81a} have listed an integrated rotation measure
of $-67\pm3$~rad~m$^{-2}$, corresponding to a rotation of about $5^{\circ}$ 
at 8~GHz.
The total intensity images at 5 and 8~GHz show a structure resembling a 
counterjet, as previously reported by \citet{Giovannini01}; indeed, our 
spectral-index map shows 
the core region itself to be optically thick, but the emission to the 
southwest of the core to be optically thin. 
Using the jet to counterjet surface 
brightness ratio, $R_j=(\frac{1+\beta~cos~\theta}{1-\beta~cos~\theta})^p$ 
\citep[Appendix~A,][]{UrryPadovani95}
at a distance of $\sim$2 mas from the core, we obtain $\beta$~cos~$\theta$=0.47 
for $R_j\sim22$ and $p (= 2 + \alpha)$ = 3. 
{ We can also estimate the doppler factor, $\delta$, using the 
ratio of the observed radio core prominence parameter, $R_c$, 
to the intrinsic radio core prominence, $R_c^{int}$, 
\citep[$\delta=(\frac{R_c}{R_c^{int}})^{1/p}$,][]{UrryPadovani95},
keeping in mind the caveat that $R_c$ is a statistical measure of orientation. 
Assuming $R_c^{int}$ to be 0.33 and $R_c$=0.05 for 3C66B 
\citep[see][]{Kharb04}, the above derived $\beta$~cos~$\theta$ gives a jet 
inclination angle of $\sim60\degr$ for 3C66B.}

{\it 3C78:}
Among the three FRI radio galaxies considered here, 3C78 shows the most 
extensive polarization along its radio jet. We detect polarization in the
jet at $\sim$3 and 6~mas from the VLBI core at 5, 8 and 15~GHz.
The 8~GHz image looks very similar to those from our epoch-2002
observations \citep{Kharb05} with the exception of a polarized core.
We did not detect core polarization at any frequency in our present data.
{ This is likely to be due to variability in the core, since our present
image sensitivity is an improvement by a factor of $\sim$1.5, over the
2002 observations.}
The total intensity 
images at 5 and 8~GHz show evidence for a counterjet, and this appears
to be borne out by our spectral-index map, which shows the core
to be optically thick, but the emission to the southwest of
the core to be optically thin. 
The jet to counterjet ratio, $R_j$ varies from $\sim32$ at $\sim1$~mas to
$\sim$22 at $\sim$2.5~mas. This results in $\beta$~cos~$\theta$ varying from
0.52 to 0.47. 
{ Again assuming $R_c^{int}$=0.33 and $R_c$=0.38 for 3C78 
\citep[see][]{Kharb04}, the derived $\beta$~cos~$\theta$ results in a jet 
inclination angle of $\sim50\degr$ for 3C78.}

\begin{figure}[h]
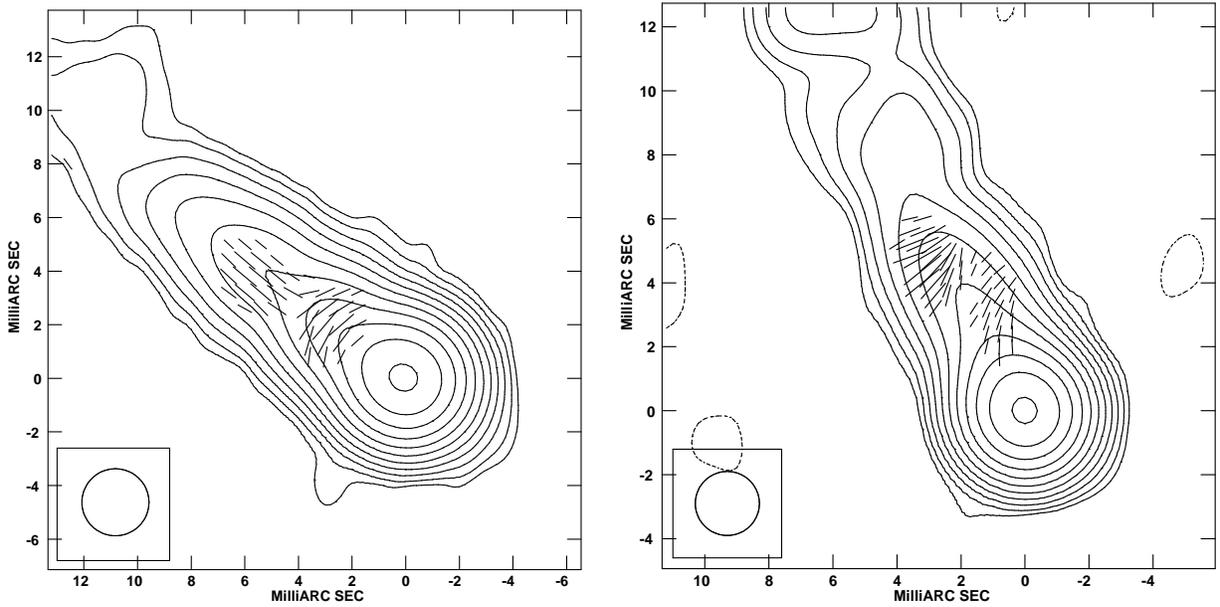

\centering{
\includegraphics[width=8.0cm]{f9a.ps}
\includegraphics[width=8.3cm]{f9b.ps}}
\caption{The polarization maps with 5~GHz total intensity contours of 
3C78 (left) and 3C264 (right) with the effects of {\it RM} removed. 
Circular beams of 2.5 and 2.0 mas were used for the 3C78 and 3C264 maps, 
respectively. $\chi$ vectors for 3C78 and 3C264: 
1 mas = 0.6 and 0.4 mJy beam$^{-1}$, respectively.
The intrinsic EVPA for the jet in 3C78 changes from nearly longitudinal 
to nearly transverse to the VLBI jet direction, while it remains roughly 
transverse throughout in 3C264.}
\label{fig:intr}
\end{figure}

3C78 is the only one of the three radio galaxies for which the rotation 
measure could be estimated using all the three frequencies.
The rotation measure in 3C78 ranges from about +200 rad~m$^{-2}$ 
to $-$400 rad~m$^{-2}$. We observe an {\it RM} difference of $>300$~rad~m$^{-2}$
transverse to the inner VLBI jet over a region that is 
$\sim$5~mas (two beam-sizes) across (Fig.~\ref{fig:3c78rm}). 
As the EVPA rotates through almost $90\degr$ 
at the lower edge of the jet (Fig.~\ref{fig:3c78rm}), 
the low energy thermal electrons that produce the rotation cannot
be completely mixed in with the synchrotron-emitting jet volume 
\citep{Burn66,CioffiJones80}, and are therefore external to it.
We discuss this further in \S 5.
Figure~\ref{fig:intr} shows the polarization map of 3C78 after correcting for
the {\it RM}.
Since these regions are optically thin, the jet magnetic field is
inferred to be aligned to the jet direction, in the region of the {\it RM} 
gradient. 
However, the magnetic field becomes orthogonal to the jet flow in
the next jet component, and appears to remain orthogonal substantially
further from the core (Fig.~\ref{fig:3c78}, top left).

{\it 3C264:}
Our observations reveal polarized emission at 5 and 8~GHz in two regions in 
the jet $\sim$1 and 4~mas away from the core, similar to that observed
by \citet{Kharb05}. We could obtain an {\it RM} map using only the 5 and 
8~GHz maps. 
Therefore, the polarization angles are subject to $\pm$n$\pi$ ambituities,
which are impossible to resolve.
The two-frequency {\it RM} across the 
inner-jet varies from $\sim+$250 rad~m$^{-2}$ to $\sim-$300 rad~m$^{-2}$.
Note that an ambiguity of $\pm180\degr$ between 5 and 8~GHz could 
result in an {\it RM} change of $\gtrsim\pm2500$ rad~m$^{-2}$. However, 
we find that the two-frequency {\it RM} in 3C264 is of the same 
order as the three-frequency {\it RM} in 3C78 and other radio
galaxy jets (Table~\ref{fr1fr2}).
Figure~\ref{fig:intr} shows the polarization map of 3C264 with the 
effects of the tentative two-frequency {\it RM} removed.
If the two-frequency {\it RM} values are correct, the jet magnetic field is
roughly aligned with the VLBI jet direction.

\begin{deluxetable}{ccccccccccccc}
\tablecaption{Observed Parameters}
\tablewidth{0pt}
\tablehead{
\colhead{Source}&\colhead{$z$}&\colhead{Scale}&\colhead{Freq}&\colhead{$I_{peak}$}& 
\colhead{$I_{tot}$}&\colhead{$I_{rms}$}&\colhead{$P_{peak}$}&\colhead{$P_{rms}$} \\
\colhead{}  & \colhead{}  & \colhead{pc/mas}&\colhead{GHz}&\colhead{mJy/bm}& 
\colhead{mJy}&\colhead{$\mu$Jy/bm}&\colhead{$\mu$Jy/bm}&\colhead{$\mu$Jy/bm}}
\startdata
3C66B& 0.0212 & 0.424 &5.0&113.9&186.2&43.4&...&34.3\\
     &        &       &8.4&110.1&193.2&46.9&636&37.5\\
     &        &       &15.3&97.4&160.9&87.8&...&66.6\\
3C78 & 0.0286 & 0.567 &5.0&342.1&547.8&70.4&846&40.4\\
     &        &       &8.4&330.9&548.1&64.7&993&39.8\\
     &        &       &15.3&242.7&416.5&183.7&720&97.1\\
3C264& 0.0217 & 0.433 &5.0&144.4&191.6&47.5&578&39.8\\
     &        &       &8.4&139.7&186.8&51.8&364&39.4\\
     &        &       &15.3&106.3&142.1&96.3&...&70.2\\
\enddata
\tablecomments{
Col.1: Source name. Col.2: Redshift. Col.3: Parsecs corresponding
to 1 mas in source. Col.4: Observing frequency in GHz. Col.5: Peak surface 
brightness in total intensity ($I$) map.
Col.6: Total VLBI flux density. Col.7: $RMS$ noise in $I$ map.
Col.8: Peak surface brightness in polarized intensity ($P$) map.
Col.9: $RMS$ noise in $P$ map.}
\label{tabobsparam}
\end{deluxetable}

\subsection{Parsec-scale {\it RM} in different AGN classes}

\begin{deluxetable}{lllllllll}
\tablecaption{Radio galaxies with measured parsec-scale rotation measures}
\tablewidth{0pt}
\tablehead{
\colhead{Source}&\colhead{FR}&\colhead{z}&\colhead{Int. RM}&\colhead{RM}
&\colhead{Comment}&\colhead{Ref}\\
\colhead{Name}  &\colhead{Type}&\colhead{}&\colhead{rad~m$^{-2}$}&
\colhead{rad~m$^{-2}$}&\colhead{}&\colhead{}}
\startdata
3C78 & I  & 0.0286 &+14&$-400,+200$& in jet at $r$= 1.7, 4.3 pc & 1\\
3C264$^a$& I & 0.0217 &+16&$-250,+150$& in jet at $r$= 2.3, 1.9 pc  & 1\\
3C84 & I  & 0.0175 &.....$^b$&+7000& in jet at $r$= 5 pc & 2 \\
M87  & I  & 0.0043 &+872&$-4500$ & in jet at $r$= 2 pc  & 3 \\
3C111& II & 0.0485 &$-$19&$-750,-200$ & in jet at $r$= 2.8, 4.7 pc & 3 \\
3C120& II & 0.0330 &$-$3&$+100^{\dagger}$& in jet at $r$= 1 pc& 3 \\
     &    &        &    &$+1500^{\ddagger},+5000$& in jet at $r$= 1, 2 pc&4\\
3C166& II & 0.2449 &+101$^c$&$-2300^\ast,-310^\ast$&in jet at $r<$ 3.8, 7.6 pc&5\\ 
\enddata
\tablecomments{Comparison of FRI and FRII radio galaxies with
parsec-scale {\it RM} measurements available at present.
$^a$ Parsec-scale {\it RM} was obtained using only two frequencies, 
and is subject to the $\pm$n$\pi$ ambiguity.
Col.4 lists the integrated {\it RM} from \citet{Simard-Normandin81a}.
$^b$ Unpolarized in \citet{Simard-Normandin81b}.
$^c$ Integrated {\it RM} was estimated from values listed in
\citet{Simard-Normandin81b}.
Col.5 lists the average {\it RM} at a projected distance $r$ from the core. 
${\dagger}$ Core {\it RM} of $+$2080 rad~m$^{-2}$ was detected in 3C120.
${\ddagger}$ \citet{Gomez08} have suggested a time variability in the jet 
{\it RM} of 3C120.
$\ast$ Rest-frame rotation measure, obtained
by multiplying $RM$ by $(1+z)^2$=1.55 for 3C166.
References: 1 -- Present paper. 
2 -- \citet{TaylorGugliucci06}
3 -- \citet{ZavalaTaylor02}.
4 -- \citet{Gomez08}.
5 -- \citet{Taylor01}.}
\label{fr1fr2}
\end{deluxetable}

Prior to this study, parsec-scale rotation measures had been obtained for only 
the broad-line FRII radio galaxies 3C111 and 3C120 
\citep{ZavalaTaylor02}, the narrow-line FRII radio galaxy 3C166
\citep{Taylor01} and the FRI radio galaxies M87 
\citep{Junor01,ZavalaTaylor02} and 3C84 \citep{TaylorGugliucci06}. 
It is worth pointing out that 3C120 is a peculiar radio source. Although it has
sometimes been classified as an FRI \citep[e.g.,][]{OwenLaing89}, it possesses a
quasar-like prominent broad-line spectrum \citep{Tadhunter93}. Akin to FRII sources,
it lacks counterjet emission and exhibits superluminal jet motion 
on scales of a few hundred parsecs \citep{Walker87}. 
Table~\ref{fr1fr2} lists all available VLBI {\it RM} estimates for the radio
galaxies. 
A non-uniform parsec-scale rotation measure distribution has also been
observed in 3C166 \citep{Taylor01}.
A recent paper by \citet{Gomez08} demonstrates a clear {\it RM} gradient
across the jet in 3C120.

The absolute rotation measure estimates for the parsec-scale jets of quasars and BL~Lacs 
lie typically in the range of 100$-$500 rad~m$^{-2}$ \citep{ZavalaTaylor04}.
This is interesting because the absolute parsec-scale jet {\it RM} values of 
the FRI and FRII radio galaxies studied so far, although few, fall in the 
same range. Further, the galaxies studied so far seem to reside in 
widely varying environments, from the centre of clusters (e.g., 3C264) to 
being relatively isolated (e.g., 3C78). Therefore the similarity in 
the {\it RM} values across the parsec-scale jets of different classes of AGN, 
suggests that the Faraday rotating medium is perhaps intrinsic and closely
associated with the jet itself.

\section{The Faraday rotating medium}
\subsection{Galaxy ISM}
It has been clear from the earliest VLBI Faraday rotation measurements
\citep[e.g.,][]{Taylor98,Nan99,Nan00,GabuzdaPushkarev01,Reynolds01}
that the parsec-scale {\it RM} distributions are usually non-uniform, indicating 
that there must be a substantial component of Faraday rotation that is 
intrinsic to the AGNs, rather than arising in our own Galaxy 
or on larger scales in 
the host galaxy of the AGN \citep[e.g.,][]{Taylor98,Udomprasert97}. 
The {\it RM} varies on scales of a few parsec in 3C78 and 3C264.
The {\it RM} gradient across the VLBI jet of 3C78 is observed over a $\sim5$~mas
($\sim$ 3~pc) region. Therefore it is difficult to assign the role of the 
Faraday rotating medium to the ISM in our Galaxy or the radio host galaxy.
Recent X-ray observations of elliptical galaxies \citep[see][]{Mathews03} 
suggest that the electron density of the hot gas in the centers is 
typically $\sim0.1~$cm$^{-3}$ and falls off with radius ($r$) as
$n_e\propto r^{-1.25\pm0.25}$.
Assuming that the {\it RM} is produced in the hot gas prevalent on scales of a 
100 pc, it would produce an {\it RM} $>$40,000 rad~m$^{-2}$ for a 
uniform magnetic field, where the line of sight magnetic field strength is 
equal to the `equipartition' magnetic field (say $\sim$5~mG, see \S5.2).
This is clearly not observed. However, if the magnetic field was not uniform 
and there were `N' reversals in the field direction along the line of sight, 
then the observed {\it RM} of $\sim$200~rad~m$^{-2}$ would require N$>40,000$. 
Such a 
large number of field reversals would however result in a net depolarization 
on parsec-scales, which does not seem to be the case, as discussed below. 

\subsection{Radio Jet and Surrounding Medium}
Considering the medium surrounding the parsec-scale jet: the broad-line region 
(BLR) has been inferred to have extents typically less than a parsec 
\citep{UrryPadovani95}. BLR clouds could therefore be ruled out as a potential 
Faraday rotating medium for the jet region lying a couple of parsecs away from 
the core. The narrow-line region (NLR) could however, extend to hundreds of 
parsecs and have a large covering factor 
\citep[$\sim30-50\%$,][]{NetzerLaor93,RowanRobinson95}.
The NLR clouds could therefore in principle contribute to the observed rotation 
measures. However, the short timescale variability observed in the parsec-scale 
{\it RM} suggests that NLR clouds or the intercloud gas cannot
be the primary Faraday rotating medium, but rather this medium
is closely linked with the relativistic jet itself 
\citep{AsadaInoue08,ZavalaTaylor05}.

\begin{deluxetable}{ccccccccccccc}
\tablecaption{Derived Parameters}
\tablewidth{0pt}
\tablehead{
\colhead{Source} & \colhead{$L_{rad}$}  & \colhead{$E_{min}$} & 
\colhead{$B_{min}$} & \colhead{$P_{min}$} & \colhead{$n_e$} \\
\colhead{} & \colhead{erg/s} & \colhead{erg} & \colhead{mG} & 
\colhead{dynes/cm$^2$} & \colhead{cm$^{-3}$} }
\startdata
3C66B& 3.9E+40 & 6.9E+51 & 8.1 & 6.1E-06 & ... \\
3C78 & 2.1E+41 & 4.6E+52 & 7.1 & 4.7E-06 & 0.03\\
3C264& 4.3E+40 & 1.2E+52 & 5.7 & 3.0E-06 & 0.04\\
\enddata
\tablecomments{
Col.1: Source name. Col.2: The averaged total radio luminosity. 
Col.3 \& 4: particle (electrons and protons) energy and magnetic field 
strength at the minimum pressure P$_{min}$, obtained assuming the 
`equipartition' condition. Col.5: Minimum Pressure. Col.6: Electron density
derived for an {\it RM} of 200~rad~m$^{-2}$ using the `equipartition'
$B$-field strength and path length $dl$=1 pc.}
\label{tabparam}
\end{deluxetable}

We created depolarization maps for 3C78 and 3C264 using the relation 
$DP=\frac{m_l}{m_h}$, where $m_l$, $m_h$ are the degree of polarization 
maps at the lower and higher frequencies, respectively. 
We made $DP$ maps for 3C78 between 5 and 8 GHz (polarized intensity maps 
were blanked below 10$\sigma$) and between 
8 and 15 GHz (polarized intensity maps were blanked below 3$\sigma$). 
The mean 5-8 GHz and 8-15 GHz depolarization values in 3C78 in the 
region $\sim3$~mas away from core were $DP^5_8= 0.98 \pm 0.11$ and
$DP^8_{15}= 0.79 \pm 0.15$, while in the region $\sim6$~mas away they were
$DP^5_8= 0.60 \pm 0.07$ and $DP^8_{15}= 0.64 \pm 0.09$. 
We note that the $DP^8_{15}$ value in the region closer to
the core ($\sim3$~mas) was obtained over a much smaller region than $DP^5_8$.
Overall these large $DP$ values suggest that there was little 
depolarization at longer wavelengths in 3C78.
The depolarization map for 3C264 was made using the degree of 
polarization maps at 5 and 8 GHz (polarized intensity maps were blanked 
below 5$\sigma$). The mean depolarization was $DP^5_8= 0.44 \pm 0.11$
around $\sim1$~mas away from the core. 

We noted in \S4 that in 3C78, the polarization angle seems to rotate
by almost $90\degr$ at the lower edge of the jet (Fig.~\ref{fig:3c78rm}) 
which indicates that the Faraday rotating medium is not mixed with the main 
synchrotron-emitting jet. Furthermore, the low depolarization suggested by the 
$DP$ estimates and the strong dependence of polarization angles on the 
square of the observing wavelength, supports the picture of a Faraday screen 
with a relatively constant Faraday depth. Since the hot gas observed in 
X-rays in fact requires a highly tangled magnetic field
to produce the observed RM, which does not seem to be borne out
by the depolarization measurements, we suggest that an optically thin
layer surrounding the jet that has sufficient thermal
electrons mixed in with any synchrotron-emitting electrons could be
a good candidate for the Faraday screen. 

Sheaths around jets have been suggested to be present in both FRI and FRII 
jets on both kiloparsec- \citep{Hardcastle96,Katz-Stone99} and parsec-scales 
\citep{Attridge99,Kharb08}. Either shearing due to the jet interaction with
the ambient medium or the presence of an intrinsic toroidal or helical 
magnetic field could result in these sheaths having ordered 
magnetic fields \citep{Laing81,Lyutikov05}. Although a sheath resulting from 
jet entrainment has often been proposed for FRI jets 
\citep{DeYoung84,Bicknell96}, sheaths could also result from the diffusion 
of particles out of the beam into the external medium \citep{ODeaOwen87}, or 
from material associated with the inner part of a disk wind 
\citep{BlandfordLevinson95,HanaszSol96}. The fact that we observe similar 
parsec-scale jet rotation measures in FRIs and FRIIs, where the levels of 
entrainment are expected to differ \citep[e.g.,][]{DeYoung93}, appears to 
provide support to the latter two models, under the assumption that
jet-like magnetic field lines also thread the sheath region.

The most natural interpretation for the {\it RM} gradient observed in 3C78 
is that it is due to a helical or toroidal jet magnetic field 
\citep{Blandford93,Laing96}. {This is also consistent with the 
increase in the degree of polarization towards the edge of the jet
(Fig~\ref{fig:fpol}).}
The asymmetry in the {\it RM} values is likely 
to be the combined effect of the jet orientation and the pitch angle of the 
field \citep[for example, see][]{Asada02}. Similar {\it RM} gradients have 
been observed in a number of other AGNs, including several 
BL~Lac objects 
\citep{Asada02,GabuzdaMurray04,ZavalaTaylor05,Gabuzda08}. 
Thus, the {\it RM} gradient together with the large rotation 
in the polarization angles and the low depolarization supports the idea that 
the Faraday rotating medium is a `sheath' around the parsec-scale jet, which is 
threaded by a helical magnetic field that is intrinsically associated with 
the jet.

The electron density in the medium producing the Faraday rotation can be 
derived under the assumption that the condition of `equipartition' between 
magnetic field and electron energy density \citep{Burbidge59} holds in the 
radio jet \citep[e.g.,][]{Feigelson95,Harris02}. Using the 5~GHz jet flux 
density and volume we obtained the magnetic field strength, electron energy 
density and pressure (Table~\ref{tabparam}) under the `equipartition' condition 
\citep[see][]{ODeaOwen87}. For these calculations, we assumed the ratio of 
the relativistic proton to relativistic electron energy to be unity.
The radio spectrum was assumed to extend from 10~MHz to 100~GHz.
An average spectral index of 0.8 and a volume filling factor 
of unity was adopted for the jet. 
The constant $C_{12}$ \citep{Pacholczyk70} which depends on the
spectral index and frequency cutoffs was taken to be $6.5 \times 10^7$. 

Assuming that the Faraday rotating medium was external to the 
synchrotron-emitting jet volume but close to it, with the magnetic field 
strength being similar to the jet `equipartition' $B$-field of $\sim 5$~mG
a canonical rest-frame 
rotation measure of 200~rad~m$^{-2}$ and a path length of 1 pc, we derive a 
lower limit to the electron density which is a few times $0.01~$cm$^{-3}$ for 
the medium in 3C78 and 3C264. 
If we assume that the path length is in fact the width of an outer sheath layer,
which is (say) about $10\%$ of the jet width ($\approx$5 pc, estimated from the
5~GHz maps), we obtain an electron density of $\sim0.1~$cm$^{-3}$, which is of 
the same order as the electron density of the hot gas in the galaxy centers.
Therefore if the sheath was a `mixing' layer between the jet and the
surrounding hot gas, it could certainly have the required electron
density, magnetic field strength and path length, to produce the observed {\it RM}.

{Although helical jet $B$-fields provide an attractive explanation for the
transverse {\it RM} gradient and increase in fractional polarization at the
jet edges in 3C78, alternative explanations, such as an inhomogeneous
surrounding medium (e.g., accretion disk wind) that gives rise to a local
enhancement in the {\it RM} on one side of the jet, cannot be completely ruled
out. The absence of a clear transverse {\it RM} gradient in 3C78 further from the
core may provide evidence for an inhomogeneous ambient medium, whether
or not this jet carries a helical $B$-field. The jet of 3C264 does not show any
clear gradients in our two-frequency {\it RM} map, suggesting that either this jet
does not carry a helical $B$-field, or that, if the jet does carry a helical
field, this field is distorted by interaction with the surrounding medium.
Higher-resolution low-frequency observations that are sensitive
to Faraday rotation but better resolve the jet in the transverse direction,
such as may be provided by future space VLBI projects, may help resolve these
questions.}

\subsubsection{Clues from optical observations}
The three FRIs under study have been observed with optical imaging
polarimetry with the {\it HST} at a resolution of
$\sim0.1\arcsec$ (translating to scales of hundreds of parsecs for 
these sources) by \citet{Perlman06} and \citet{Capetti07}.
\citet{Perlman06} found that 3C66B, 3C78 and 3C264 showed the simplest 
magnetic field structures among the six nearby radio galaxies studied.
Interestingly, both 3C78 and 3C264 (sources with significant parsec-scale 
polarization) exhibit an increase in the degree of polarization along the edges of 
the optical jet.
\citet{Perlman06} suggest that such a magnetic field structure could
arise due to a spine-sheath jet structure, with the sheath 
forming due to the interaction of the jet with the surrounding medium, which 
stretch the magnetic field lines due to shear. We note that a spine-sheath
structure could also simply result from a helical jet magnetic field,
with or without shearing \citep{Lyutikov05}.

A `potential' problem with the jet interaction picture could be that on 
going down from $\sim0.1\arcsec$ to mas-scales, we detect polarization along the 
centre of the jet (in 3C78 and 3C264) but not at the edges as expected from 
a `sheath'. This could either be due to insignificant interaction/entrainment 
on parsec-scales, or, 
{the existence of a helical magnetic field extending from
parsecs (where its presence is revealed by the observed transverse {\it RM} gradient)
to hundreds of parsecs (where its presence is revealed by the
observed transverse optical polarization structure).} This 
helical field could originate due to the rotation of the accretion disk plus 
jet outflow, and propagate outward with the jet throughout its length. 
Another possibility is the scenario proposed by \citet{Asada08} for the 
quasar NRAO140, $viz.,$ that the jet has a spine-sheath structure with 
helical magnetic fields threading both the jet components, but the field is 
tightly wound in the sheath (resulting in the {\it RM} gradient) and loosely 
wound in the spine (leading to aligned magnetic fields in the inner jet).
This scenario is also consistent with the findings in 3C120 and 3C166.
Interestingly, such a magnetic field geometry has been known to result in the 
matter + Poynting flux dominated jets produced in the 3-D general relativistic 
MHD simulations of \citet{DeVilliers05} and \citet{Hawley06}.
Such a magnetic field geometry could result in net depolarization at the jet 
edges which have transverse fields as opposed to the dominant spine which have
aligned fields, thereby resolving the `potential' problem.
Significant interaction/entrainment on scales of a few hundred 
parsecs, along with magnetic flux conservation could reproduce the 
tranverse $B$-field spine and longitudinal $B$-field sheath 
observed in kiloparsec-scale radio galaxy jets \citep{Canvin05,Laing06}.

Another interesting feature that emerges from optical observations is the 
presence of optical jets in the majority of radio galaxies with
detected parsec-scale rotation measures, $i.e.,$ detected parsec-scale polarization
over two or more radio frequencies, which in turn could be the result of low
depolarization between frequencies. With the exception of 3C111 and 3C166, 
all the radio galaxies listed in Table~2 have optical jets.
However, even 3C166 exhibits an optical `tail' along the radio jet 
direction \citep{deKoff96}. Both 3C111 and 3C166 
also exhibit unresolved {\it HST} nuclei which have been suggested to 
be the unresolved bases of optical jets \citep{Chiaberge00}. 
A possible connection between the presence of optical emission in the jet 
and detectable parsec-scale polarization in the radio needs to be examined further
with a greater number of sources.

However if there was such a connection, it can be easily understood.
Optical jet emission is expected to result from particle re-acceleration
and to some extent, Doppler boosting \citep[e.g.,][]{Sparks94,Bicknell96}.
Two of the three FRIIs with detected polarized emission, 
$viz.,$ 3C111 and 3C120, are broad-line radio galaxies, and
are likely to exhibit Doppler boosting in their jets {due to being oriented
at relatively small angles to line of sight. 
Furthermore, the kpc-scale radio core prominence values of the three FRIs
under study
indicate that the most polarized source, 3C78, is also the most 
core-dominant, while the least polarized source, 3C66B, is the least
core-dominant. This underscores the importance of Doppler boosting
in these parsec-scale jets.}
Particle re-acceleration could occur in the jet shear layer or sheath, as
the jet interacts with the surrounding medium, thereby producing a net 
deceleration and converting bulk kinetic energy into high energy 
optical radiation \citep{Owen89,Stawarz03}. 

\section{Summary and Conclusions}
We have observed three FRI radio galaxies, $viz.,$ 3C66B, 3C78 and 3C264, with
polarization sensitive VLBI at 5, 8 and 15 GHz. We detected polarization in
all three and obtained rotation measures
across the parsec-scale jets of 3C78 and 3C264. $RM$ was obtained using all the
three frequencies in 3C78 but only two frequencies in 3C264, which is therefore 
subject to uncertainty due to $\pm$n$\pi$ ambiguities that cannot be resolved.
To summarise,

\begin{enumerate}
\item Polarization was detected in the jets of 3C66B at 8 GHz,
in 3C78 at all the three frequencies and in 3C264 at 5 and 8 GHz.

\item The total intensity maps confirm the detection of a counterjet in 
3C66B and indicate the presence of one in 3C78.
The 5-8 GHz spectral index maps reveal an optically thick core and
an optically thin jet in all sources. The counterjet emission in 
3C66B and 3C78 is also optically thin.

\item The parsec-scale jet rotation measure in 3C78 ranges from 
$\sim$+200 rad~m$^{-2}$ to $\sim-$400 rad~m$^{-2}$. 
The core appears to be completely depolarized.
A rotation-measure gradient is observed across the jet in 3C78 in a region 
that is over two beam-sizes across. Furthermore, the degree of polarization 
increases along the edge of the jet. 
%This provides clear evidence for the presence of a helical jet magnetic field.
This strongly supports the idea of a helical jet magnetic field.

\item The polarization angles rotate by almost 90$\degr$ between the
three frequencies in 3C78. The depolarization parameter is close to unity
in the jet, suggesting low depolarization at longer wavelengths.
This suggests that the Faraday rotation is occuring in a layer
containing thermal plasma that is external to the main body of the jet.

\item After correcting for the Faraday rotation, the magnetic field in 
3C78 seems to be aligned with the jet direction close to the core, but 
becomes orthogonal further down the jet. 

\item The two-frequency parsec-scale rotation measure in the 3C264 jet varies from 
$\sim$+250 rad~m$^{-2}$ to $\sim-$300 rad~m$^{-2}$. 
The core is completely depolarized.

\item The parsec-scale rotation measures in 3C78 and 3C264 
are similar to those observed in other nearby FRI and FRII radio galaxies.
Parsec-scale $RM$ of a few hundred rad~m$^{-2}$ have also
been inferred in the jets of the Doppler-beamed BL~Lacs and quasars. 
The similar parsec-scale rotation 
measures in these AGN classes suggests an intrinsic origin, perhaps
in the jet medium itself.
\end{enumerate}

Based on the transverse {\it RM} gradient, increase in the degree of
polarization towards the edge of the jet, large rotation of the
polarization angles due to the Faraday rotation, and the small
depolarization in 3C78, we argue that a layer surrounding the jet and
carrying a helical magnetic field forms the Faraday screen. This
suggestion is supported by {\it HST} optical polarization images. The recent
work by \citet{Gomez08} on 3C120 is likewise fully consistent with
our conclusions.

The presence of optical jets in the majority of radio galaxies with
detected parsec-scale polarization and the spine-sheath polarization
structure observed in the {\it HST} polarization images of some of these
sources suggests a simple connection: a sheath layer around the jet,
produced either by jet-medium interaction or a helical magnetic field, 
is producing both. {Jet-ISM entrainment and interaction could cause
(1) mixing of thermal gas with the outer sheath layer of the jet which 
produces the rotation measure, and (2) particle acceleration which produces
the optical jet.} This connection needs to be further explored based on a 
larger number of sources.

\acknowledgments
We are grateful to the referee for a careful assessment 
of our work which has improved this paper.
We thank Matthew Lister, Maxim Lyutikov, Robert Laing and Keiichi Asada
for providing valuable insights towards issues related to this paper.
This research has made use of data from the University of Michigan Radio 
Astronomy Observatory which is supported by funds from the University of Michigan.
The National Radio Astronomy Observatory is a facility of the National Science 
Foundation operated under cooperative agreement by Associated Universities, Inc.
This research has made use of the NASA/IPAC Extragalactic Database (NED) which 
is operated by the Jet Propulsion Laboratory, California Institute of 
Technology, under contract with the National Aeronautics and Space 
Administration.


{\it Facilities:} \facility{VLBA}, \facility{EVN}.

%\bibliographystyle{mn2e}
%\bibliography{ms}

\end{document}